\documentclass[aps,prl,twocolumn,groupedaddress]{revtex4-1}

\usepackage{graphicx}
\usepackage{amsmath}
\usepackage{relsize}

\begin{document}

\title{Coulomb drag and depinning in bilinear Josephson junction arrays}

\author{Samuel A. Wilkinson}
\affiliation{Chemical and Quantum Physics, School of Science, RMIT University, Melbourne, Victoria 3001, Australia}
\author{Nicolas Vogt}
\affiliation{Chemical and Quantum Physics, School of Science, RMIT University, Melbourne, Victoria 3001, Australia}
\author{Jared H. Cole}
\affiliation{Chemical and Quantum Physics, School of Science, RMIT University, Melbourne, Victoria 3001, Australia}

\date{\today}

\begin{abstract}
Coulomb drag and depinning are electronic transport phenomena that occur in low-dimensional nanostructures. Recently, both phenomena have been reported in bilinear Josephson junction arrays. These devices provide a unique opportunity to study the interplay of Coulomb drag and depinning in a system where all relevant parameters can be controlled experimentally. We explain the Coulomb drag and depinning characteristics in the I-V curve of the bilinear Josephson junction array by adopting a quasicharge model which has previously proven useful in describing threshold phenomena in linear Josephson junction arrays. Simulations are performed for a range of coupling strengths, where numerically obtained I-V curves match well with what has been previously observed experimentally. Analytic expressions for the ratio between the active and passive currents are derived from depinning arguments. Novel phenomena are predicted at voltages higher than those for which experimental results have been reported to date.

\end{abstract}

\maketitle

\section{Introduction}
When two systems with long-ranged Coulomb interactions are placed in close proximity, applying a voltage bias to only one of these systems can produce a current in both, even when there is no direct transfer of charge carriers. This effect, known as Coulomb drag \cite{Narozhny2016}, is of fundamental interest in condensed matter physics and nano-electronics, and has been observed in a wide variety of systems, including graphene bilayers \cite{Gorbachev2012,Li2016}, 2-D electron gases \cite{Zheng1993}, edges states of fractional quantum Hall systems \cite{Orgad1996}, quantum point contacts \cite{Khrapai2007}, nanowires \cite{Glazman2006,Klesse2000} and superconducting wires and films \cite{Duan1993}.

Another important transport phenomenon of interest in many-body systems is depinning \cite{Brazovskii2003}, where transport in a driven system is inhibited by disorder `pinning' the system in metastable states until the collective pinning force is overcome at some threshold driving strength. Depinning theory is crucial to the understanding of transport in systems such as charge density waves\cite{Fukuyama1978}, magnetic domain boundaries \cite{Imry1975} and flux-line lattices in type-II superconductors \cite{Larkin1979}. It has recently been shown that the onset of conduction in linear Josephson junction arrays at low voltages is also determined by the effects of depinning \cite{Vogt2015a}. A unique system in which the interplay between depinning and Coulomb drag can be observed is the bilinear Josephson junction array.

Josephson junction arrays are ideal for studying low-dimensional electronic transport in a controllable system, as almost all parameters of the system can be tailored at will.
Linear Josephson junction arrays have already been used to study depinning in a system where relevant parameters such as the amplitude of the pinning potential can be controlled \cite{Vogt2015a}.
To study both depinning and Coulomb drag, and their interplay, we focus on bilinear Josephson junction arrays.
These systems consist of two linear chains of Josephson junctions coupled capacitively, so that the two arrays interact electrostatically but no direct transfer of charge carriers occurs between the two arrays.
Experiments on these bilinear arrays have shown Coulomb drag and current mirror behaviour \cite{Shimada2000,Shimada2012}, but the quantitative details of these effects have so far not been understood theoretically, nor modelled computationally.

Linear arrays are known to exhibit threshold behaviour when the charging energy of a single grain $E_C = (2e)^2/2C_J$ is comparable to or greater than the Josephson tunnelling energy $E_J$ \cite{Haviland1996}.
In this regime the array acts as an insulator at small voltages, with a sudden transition to conducting behaviour at a switching voltage $V_{sw}$. 
Most work to date on these systems, both experimentally and theoretically, has described the threshold behaviour in terms of the minimum energy required to inject a single charge soliton into the array\cite{Haviland1996,Fedorov2011,Homfeld2011,Hermon1996}.
While this simple picture is elegant and attractive, it fails to account for effects arising from disorder in the system, which is always present in any realistic experiment.
More recently, threshold behaviour in linear arrays was described in terms of the pinning of charge due to disorder in the array\cite{Vogt2015}. 
Such an approach has been able to achieve an excellent fit between the theoretical model and experimental data\cite{Vogt2015a}.
In this work, we extend this description to bilinear Josephson junction arrays.

Two different coupling geometries for bilinear arrays have been fabricated: straight coupling \cite{Shimada2012}, where each site is coupled to only one site on the opposite array, and slanted coupling \cite{Shimada2000}, where each site is coupled to two sites on the opposite array.
Differences in the phase diagrams between these systems have been studied theoretically \cite{Choi1998d}.
For concreteness, we will focus primarily of the case of straight coupling.
However our methods can be readily applied to slanted coupling, and we will include results for both arrays where relevant.

\begin{figure}
	\centering
	\includegraphics[width=1\linewidth]{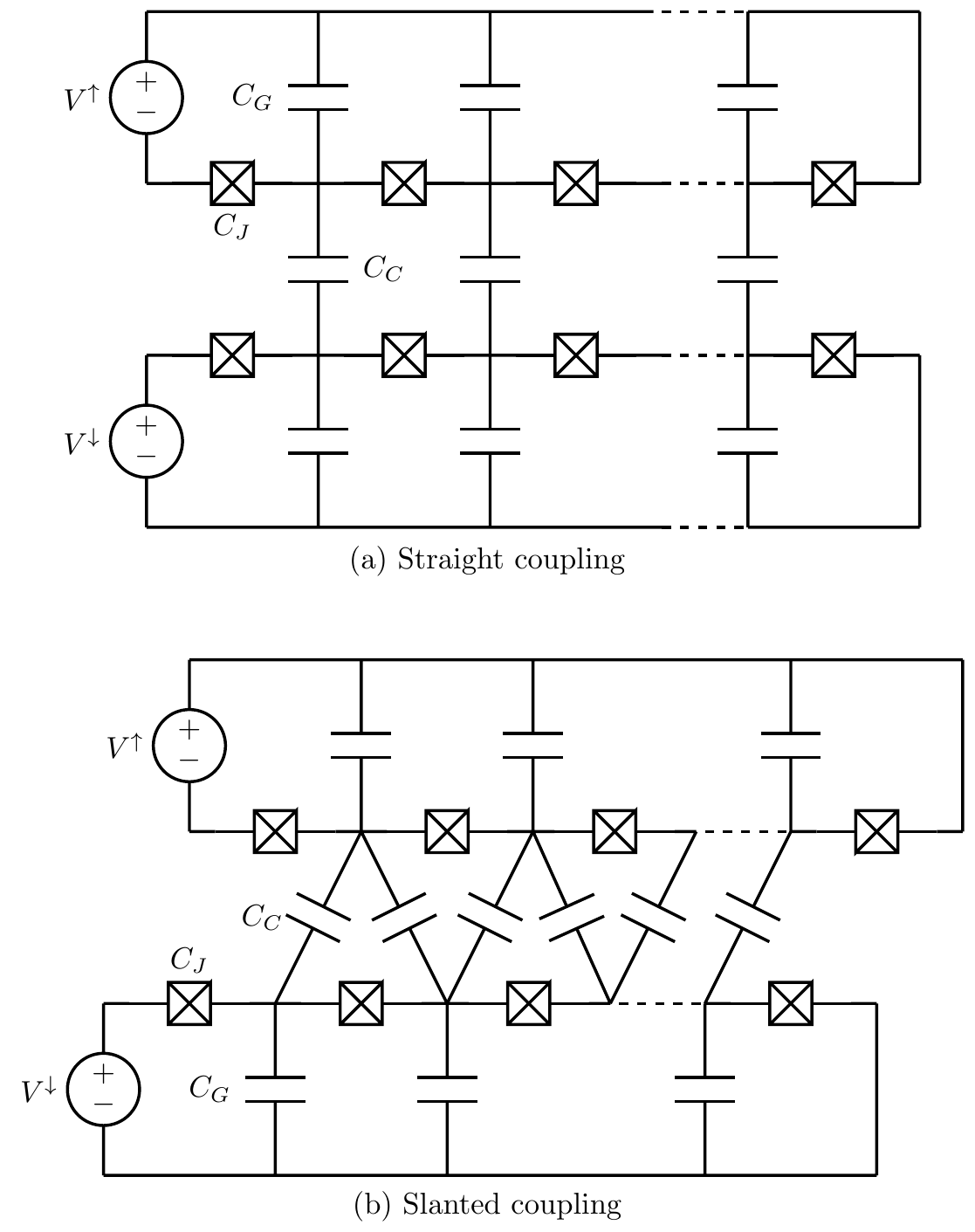}
	\caption{\label{fig:cicuit} Circuit diagram of bilinear Josephson Junction arrays with (a) straight and (b) slanted coupling. The electrostatics of these circuits are determined by the capacitance to ground $C_G$, the Josephson junction capacitance $C_J$ and the coupling capacitance $C_C$.}
\end{figure}

In the bilinear Josephson junction array with straight coupling, depicted in Fig.~\ref{fig:cicuit}, each site is coupled to two other sites on the same array via Josephson junctions, and to one site on the opposite array by a capacitance $C_C$. 
Each site also has a ground capacitance $C_G$.
There is assumed to be no tunnelling of charge carriers between the two arrays.

In a typical experiment, one of these arrays is held at a fixed voltage (assumed to be 0). 
We call this array the \textit{passive} array. 
The voltage on the \textit{active} array, however, is varied, so that an I-V curve can be determined.
Both arrays remain insulating at low voltages.
At the switching voltage $V_{sw}$ both arrays become conducting, even though no voltage is directly applied to the passive array.
The current through the active array is much larger, but the drag current in the passive array is clearly present and linearly proportional to the active bias.

Previous theoretical work on these systems has focused on their static features such as their free energy \cite{Hu1996} and phase transitions \cite{Choi1998c}, or has focused on dynamics in the quasiparticle-dominated limit \cite{Walker2013}. Here we are more concerned with the dynamical behaviour when each grain is deep in the superconducting regime, and we explicitly derive equations of motion for the charges on each island, which we use to numerically calculate I-V curves.

\section{Theoretical Model}
For clarity, we will first derive the Hamiltonian and equations of motion for the bilinear array with straight coupling.
From this it is trivial to alter our model to the case of slanted coupling.

Across each capacitor and junction in the array there will be a difference in the phase of the superconducting condensate.
We use $\phi$, $\psi$ and $\chi$ to denote the phase difference across the Josephson junctions ($C_J$), ground capacitors ($C_G$) and coupling capacitors ($C_C$) respectively (see Fig.~\ref{fig:cicuit}). 
From the circuit diagram in Fig.~\ref{fig:cicuit} (a), standard theoretical techniques \cite{Devoret1997,Zagoskin2011} can be applied to derive a Lagrangian for the system,
\begin{equation}\label{eq:Lag1}
	\mathcal{L} = \sum_{i,\zeta} \frac{C_J}{2}(\dot{\phi}_i^\zeta)^2 + \frac{C_G}{2}(\dot{\psi}_i^\zeta)^2 + \frac{C_C}{2}(\dot{\chi}_i)^2 + E_J \cos (\phi_i^\zeta)
\end{equation}
where the index $i$ runs over different sites in an array, and the index $\zeta$ runs over the two arrays.

Each phase in the Lagrangian has a conjugate charge variable
\begin{equation}
	q_i^{\phi,\zeta} = C_J\dot{\phi}_i^\zeta ;
	\qquad 
	q_i^{\psi,\zeta} = C_G\dot{\psi}_i^\zeta ;
	\qquad 
	q_i^\chi = C_C\dot{\chi}_i.
\end{equation}
These charges are related to the number of Cooper pairs on a superconducting island by charge neutrality and Kirchoff's laws, which allow us to write the Hamiltonian in terms of the number of Cooper pairs
\begin{equation}
\begin{split}
	\mathcal{H} = \sum_{i,j} 4e^2 (n_i^\zeta + f_i^\zeta) \left[C^{-1}\right]_{ij}^{\zeta\zeta'} ( n_j^{\zeta'} + f_J^{\zeta'} )\\ - E_J\cos{ \phi_i^\zeta } - E_J\cos{ \phi_i^{\zeta'} }
\end{split}
\end{equation}
where $f_i$ is the charge frustration on the $i$th junction due to, e.g., trapped charges or defects, and is modelled by a random number $f_i \in (-1,1)$, which we take to be evenly distributed across the entire interval (although other disorder distributions have been considered, such as weak or Gaussian disorder\cite{Vogt2015,Walker2015}). $C_{ij}^{\zeta\zeta'}$ is the capacitance matrix which expresses Coulomb interactions between sites (and therefore depends on the particular coupling geometry). The inverse capacitance matrix includes long-range electrostatic interactions between islands. In discussing threshold phenomena in linear chains, it has proven useful to move away from this highly coupled model to a so-called quasicharge model, which includes only nearest-neighbour interactions.

To this end, we introduce cumulative quasicharge variables 
\begin{equation}
	\begin{split}
	m_i^\zeta = -\sum_{j=1}^{i-1} n_j^\zeta; \quad F_i^\zeta = \sum_{j=1}^{i-1} f_j^\zeta; \quad
	Q_i^\zeta = \sum_{j=1}^{i-1} q_j^{\psi,\zeta} + q_1^\zeta;\\ \quad X_i = \sum_{j=1}^{i-1} q_j^\chi;\quad
	n_i^\zeta = m_i^\zeta - m_{i+1}^\zeta; \quad q_i^{\psi,\zeta} = Q_{i+1}^\zeta - Q_i^\zeta.
	\end{split}
\end{equation}
The charge variables $Q$ are defined as the polarization across the junctions rather than the number of Cooper pairs on each grain.
As was shown in Ref.~\cite{Hermon1996}, the charge solitons in a Josephson junction array can not be thought of as a single Cooper pair dressed by a polarization cloud, but rather the soliton \textit{is} the polarization cloud itself.
In our model, it is this polarization that is our relevant charge variable, and not the number of Cooper pairs.

We follow the derivation given in \cite{Vogt2015,Vogt2015a} and note that these variables are constrained by requirements of charge neutrality, such that
\begin{align}
	q_i^{\phi,\zeta} &= \sum_{j=1}^{i-1} q_j^{\psi,\zeta} + q_1^\zeta - 2e(n_i^\zeta + f_i^\zeta) + (-1)^\zeta X_i\\
	&= Q_i^\zeta - 2e(m_i^\zeta + F_i^\zeta) + (-1)^\zeta X_i.
\end{align}
Using these relations we can rewrite the Lagrangian in Eq.~\ref{eq:Lag1} in terms of quasichagre variables
\begin{equation}
\begin{split}
	\mathcal{L} = \sum_{i,\zeta} \frac{ \left(Q_i^\zeta + (-1)^\zeta X_i - 2e(m_i^\zeta + F_i^\zeta) \right)^2 }{2C_J} + \frac{\left(Q_i^\zeta - Q_{i+1}^\zeta\right)^2}{2C_G}\\ + \frac{\left(X_i - X_{i+1}\right)^2}{2C_C} + E_J\cos\left(\phi_i^\zeta\right).
\end{split}
\end{equation}

The Lagrangian can be further simplified by introducing a dimensionless coupling parameter $\alpha = C_C/C_G$ and noting that, due to Kirchhoff's laws, $X_i = \alpha(Q_i^\uparrow - Q_i^\downarrow)$. For notational clarity, we will also switch to a vector representation where $\vec{Q}_i = (Q_i^\uparrow, Q_i^\downarrow)^\textrm{T}$ (and likewise for other quantities defined on both arrays), and introducing the discrete differential operator $\nabla$ such that $\nabla\vec{Q}_i = \vec{Q}_{i+1} - \vec{Q}_i$. Finally, we introduce a coupling matrix $\mathcal{M}$.

This gives us the Hamiltonian of the system,
\begin{equation} \label{eq:Ham}
\begin{split}
\mathcal{H} = &\sum_i \left\{ \frac{1}{2C_G}\nabla\vec{Q}_i^\textrm{T}\mathcal{M}
\nabla\vec{Q}_i \right.  \\ &+ \left. \frac{1}{2C_J}\left[ \mathcal{M}\vec{Q}_i + 2e\left(\vec{m}_i  + \vec{F}_i \right) \right]^2
- E_J\cos\left(\vec{\phi}_i\right)  \right\}
\end{split}
\end{equation}
\begin{equation}
\mathcal{M} = \begin{pmatrix}
1 + \alpha & -\alpha\\
-\alpha & 1 + \alpha
\end{pmatrix},
\end{equation}
where $\alpha = C_C/C_G$.

We now assume that the $Q$ are slow-changing parameters compared with $m$ and $\phi$, which is always the case when there is a sufficiently large inductance (in \cite{Vogt2015a} it was shown that the Bloch inductance of the array is sufficient so long as the system is driven adiabatically so as to avoid Landau-Zener transitions). In this limit, we may separate the time-scales of evolution of these parameters and apply a Born-Oppenheimer approximation. We take the portion of the Hamiltonian which depends on $m_i$ and $\phi_i$
\begin{equation}
H_Q(m_i,\phi_i) = \sum_i \dfrac{1}{2C_J}\left[ \mathcal{M}\vec{Q}_i + 2e\left(\vec{m}_i + \vec{F}_i \right) \right]^2 - \cos(\vec{\phi}_i)
\end{equation}
and take $Q$ to be a constant, classical parameter. In this Hamiltonian, the coupling matrix $\mathcal{M}$ only acts on the classical parameter $Q$, and not on the quantum operators $m_i$ and $\phi_i$. Therefore, the Hamiltonian separates into the sum of single-site Hamiltonians, which can be easily diagonalized numerically. Diagonalizing $H_Q$ for various values of $Q$, and taking only the lowest energy band, gives us an effective potential $E_Q(Q^\zeta,F^\zeta) = E_Q(Q^\uparrow,F^\uparrow) + E_Q(Q^\downarrow, F^\downarrow)$. The form of this potential is equivalent to the characteristic value of Mathieu's equation \cite{McLachlan1947} $a(\nu^\zeta)$ with argument $\nu^\zeta = (1+\alpha)Q^\zeta - \alpha Q^{\zeta'} + 2eF^\zeta$, where $\zeta'$ simply indicates the array opposite to $\zeta$. This is a $2e$-periodic function which reduces to $a(\nu) \sim \cos(\nu)$ in the limit $E_J \gg E_C$. Applying this approximation, we get an effective semi-classical Hamiltonian
\begin{equation}
H = \sum_i \frac{1}{2C_G} \nabla\vec{Q}_i^\textrm{T} \mathcal{M} \nabla\vec{Q}_i + E_Q(\mathcal{M}\vec{Q}_i + 2e\vec{F}).
\end{equation}
In this Hamiltonian, there exists only nearest-neighbour coupling between islands in the array. Coupling between arrays is mediated by the matrix $\mathcal{M}$, and charge frustration in the array enters in the form of the disordered periodic potential $E_Q$.

In this model $E_Q$ is responsible for pinning the system and preventing charge transport in the arrays.
The presence of $\mathcal{M}$ in the argument of the potential indicates that this pinning potential is highly coupled.
To understand the depinning of each of the arrays separately, we introduce new charge variables which are decoupled in the pinning potential
\begin{equation}
	\Upsilon^\uparrow = (1 + \alpha)Q^\uparrow - \alpha Q^\downarrow, \qquad \Upsilon^\downarrow = (1 + \alpha)Q^\downarrow - \alpha Q^\uparrow
\end{equation}
which is the same as ``rotating" the vectors $\vec{Q}_i$ by the matrix $\mathcal{M}$.
In this rotated frame, the Hamiltonian is
\begin{align} \label{eq:HamRot}
	H = \sum_i \dfrac{1}{2C_G}\nabla\vec{\Upsilon}_i^\textrm{T}\mathcal{M}^{-1}\nabla\vec{\Upsilon}_i + E_Q(\vec{\Upsilon}_i + 2e\vec{F}_i).
\end{align}
Coupling between the two arrays occurs only in the charging term, not in the pinning potential.
The theory of depinning would thus lead us to expect that each array in this rotated frame has a separate pinning potential, and therefore a separate depinning transition.
If a voltage were to be applied to only one array in this frame, only one array will depin, and there will be no drag current.
However, when a voltage is applied only to the active array in the unrotated frame (as is the case experimentally), there is an effective voltage on each array in the rotated frame, given by
\begin{equation} \label{eq:VoltRot}
	\begin{pmatrix}
		V_\textrm{rot}^\uparrow\\ V_\textrm{rot}^\downarrow
	\end{pmatrix}
	= \mathcal{M}\begin{pmatrix}
		V_\textrm{lab}^\uparrow \\ 0
	\end{pmatrix}
	= \begin{pmatrix}
		(1+\alpha)V_\textrm{lab}^\uparrow \\ -\alpha V_\textrm{lab}^\uparrow
	\end{pmatrix}
\end{equation}
where the subscripts correspond to the voltage applied in the rotated frame and the physical laboratory frame.
The voltage applied to the upper array in the rotated frame is larger than that applied to the lower array, so we expect this array to depin first.
At this point, there will be a current flowing through the upper rotated array, but not the lower rotated array.
This regime corresponds to the point in the lab frame where both arrays are conducting.
The lab frame currents will be given by
\begin{equation}
	\begin{pmatrix}
		I_\textrm{lab}^\uparrow\\
		I_\textrm{lab}^\downarrow
	\end{pmatrix}
	= \mathcal{M}^{-1}\begin{pmatrix}
		I_\textrm{rot}^\uparrow\\
		0
	\end{pmatrix}
\end{equation}
which allows us to predict the ratio between the active and passive currents in the lab frame as
\begin{equation}
	\frac{I_\textrm{lab}^\uparrow}{I_\textrm{lab}^\downarrow} = \frac{1 + \alpha}{\alpha}.
\end{equation}

As we increase the voltage in the lab frame, the voltage on the lower array in the rotated frame will increase until it too reaches its threshold.
As can be seen in Eq.~\ref{eq:VoltRot}, the effective voltage felt by the lower array will be negative, so that in the rotated frame the two currents will flow in opposite directions.
In the rotated frame, the ratio between currents is
\begin{equation}\label{eq:IRot}
	\frac{I_\textrm{rot}^\uparrow}{I_\textrm{rot}^\downarrow} = -\frac{1+\alpha}{\alpha}.
\end{equation}
Switching back to the lab frame, the current on the passive array is 
\begin{equation}
	I_\textrm{lab}^\downarrow = \frac{\alpha}{1 + 2\alpha}I^\uparrow_\textrm{rot} + \frac{1 + \alpha}{1 + 2\alpha}I_\textrm{rot}^\downarrow
\end{equation}
which, upon using Eq.~\ref{eq:IRot} to express this in terms of only one of the rotated currents, gives us $I_\textrm{lab}^\downarrow = 0$.

Here we have seen a separation of the pinning of each array. When the first threshold is reached, the active array depins and we enter the Coulomb drag regime. As voltage is increased beyond this first threshold, we predict that there will exist a second threshold voltage, where the passive array will depin. After the second threshold, Coulomb drag behaviour will cease and the passive current will drop to 0 (provided there is no voltage directly applied to the passive array).

\begin{figure}
	\centering
	\includegraphics[width=1\linewidth]{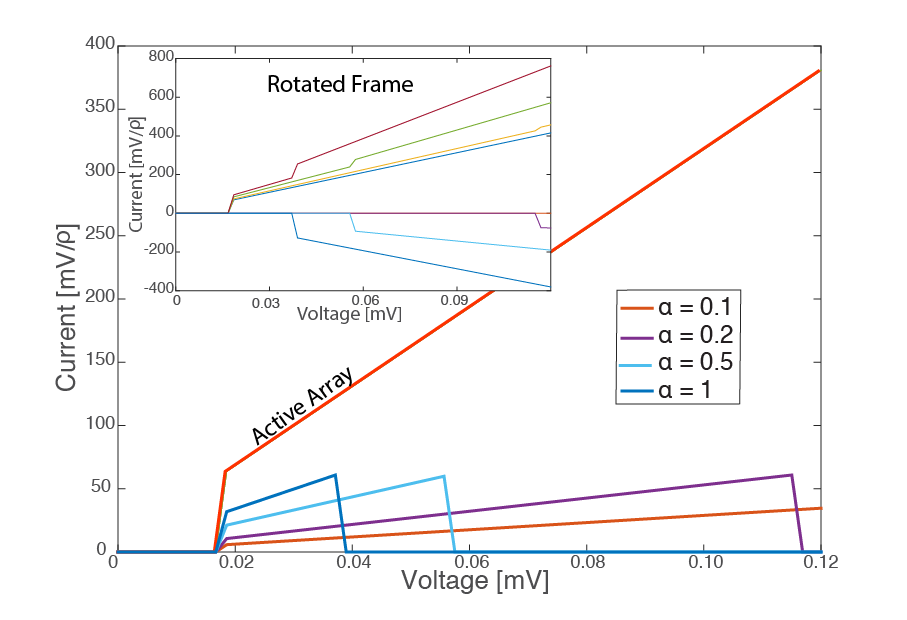}
	\caption{\label{fig:IVCurves} I-V curves for the bilinear Josephson junction array with straight coupling calculated numerically for using 20 sites per array, with $E_J/E_C = 0.5$, $C_J = 5.75 fF$, $C_G = 0.23 fF$, $\Lambda = 5$, and various values of the coupling constant $\alpha$. These values were chosen for numerical speed and convenience. Since a phenomenological resistance is used, the units of current are arbitrary and only thresholds and ratios between currents are qualitatively correct. The same I-V curves are shown in the rotated frame in an insert. Note that in the rotated frame the magnitude of both currents depends on the coupling strength. }
\end{figure}

From the Hamiltonian in Eq.~\ref{eq:HamRot} we can derive equations of motion for the variable $\vec{Q}_i$,
\begin{equation}
\frac{1}{C_G}\mathcal{M}\nabla\vec{Q}_i + V_Q(\mathcal{M}\vec{Q}_i + \vec{F}_i) + \rho\dot{\vec{Q}}_i + L\ddot{\vec{Q}}_i = 0,
\end{equation}
which in the rotated frame becomes
\begin{equation} \label{eq:OfMotion}
\frac{1}{C_G}\mathcal{M}^{-1}\nabla\vec{\Upsilon}_i + V_Q(\vec{\Upsilon}_i + \vec{F}_i) + \rho\mathcal{M}^{-1}\dot{\vec{\Upsilon}}_i + L\mathcal{M}^{-1}\ddot{\vec{\Upsilon}}_i = 0
\end{equation}

where $V_Q(Q) = \partial E_Q(Q)/ \partial Q$, $\rho$ is the resistance per site of the array and $L$ is the inductance. 
$\rho$ and $l$ are phenomenological and have been including to ensure numerical convergence (it should be noted that $\rho$ is not necessarily related to the normal-state resistance $R$, but is rather the sub-gap resistance that is observed even in the Cooper-pair limit \cite{Agren2001}).
Since these terms couple to $\dot{Q}$ and $\ddot{Q}$ respectively, these terms do not influence the location of the threshold voltages.
Because $\rho$ and $L$ are strictly phenomenological, their values are chosen for numerical convenience.
This model is therefore not able to quantitatively predict the absolute magnitude of the currents in the system.
Solving these equations of motion should, however, correctly return qualitative structure of the I-V curve, as well as quantities independent of $L$ and $\rho$, such as the ratio between currents $I^\uparrow/I^\downarrow$.

Many of these results are readily extended to the case of slanted coupling.
One must replace the 2-component quasicharge vector $\vec{Q}_i = (Q_i^\uparrow, Q_i^\downarrow)^\textrm{T}$ with the $2N$-component vector $(Q_1^\uparrow, Q_1^\downarrow, \dots Q_k^\uparrow, Q_k^\downarrow, \dots Q_N^\uparrow, Q_N^\downarrow)^\textrm{T}$, which we shall denote by $\vec{Q}$ as it should always be clear from context whether we are referring to the 2-component vector or the $2N$-component vector.
Then the $2\times 2$ coupling matrix $\mathcal{M}$ is replaces by a $2N\times 2N$ matrix
\begin{equation}
\mathcal{M} = \begin{pmatrix}
1+\alpha & -\alpha & 0 &\dots &&\\
-\alpha & 1 + 2\alpha & -\alpha &&&\\
0 & \ddots & \ddots & \ddots &&\\
\vdots && -\alpha & 1 + 2\alpha & -\alpha\\
&&& -\alpha & 1 + \alpha
\end{pmatrix}.
\end{equation}
This leads to a Hamiltonian
\begin{equation}
\begin{split}
\mathcal{H} = &\frac{1}{2C_G}\nabla \vec{Q}^\textrm{T} \mathcal{M} \nabla\vec{Q} + E_Q(\mathcal{M}\vec{Q})\\
= &\frac{1}{2C_G}\nabla\vec{\Upsilon}^\textrm{T} \mathcal{M}^{-1} \nabla\vec{\Upsilon} + E_Q(\vec{\Upsilon})
\end{split},
\end{equation}
where, as with straight coupling, $\vec{\Upsilon} = \mathcal{M}\vec{Q}$.
This gives us an equation of motion identical to Eq.~\ref{eq:OfMotion}, but with the slanted definitions of $\vec{Q}$, $\vec{\Upsilon}$ and $\mathcal{M}$

The form of the coupling matrix for the slanted array is the same as that of the capacitance matrix for a linear Josephson junction array (with the identifications $C_J \rightarrow \alpha$ and $C_G \rightarrow 1$), and so it can be inverted analytically using the same methods \cite{Hu1996,Cole2014}.

Much of the analysis of the array with straight coupling depended on the fact that the Hamiltonian separates neatly into a sum of two-site terms.
This is not the case with slanted coupling, and we are not able to obtain the same analytical results.

\section{Results of numerical simulations}
The equation of motion Eq.~\ref{eq:OfMotion} enables us to numerically determine I-V curves for the bilinear array with either slanted or straight coupling.
In Fig.~\ref{fig:IVCurves} we present such I-V curves for arrays with straight coupling calculated at a variety of different couplings $\alpha$ using parameters given in the caption to figure~\ref{fig:IVCurves}, which were chosen for numerical speed and convenience.
The existence of two distinct pinning thresholds can clearly be seen, and the predicted current ratios hold.
The qualitative form of the I-V curves at low voltages (i.e. before the second threshold) are in good agreement with the published experimental results of Ref.~\cite{Shimada2012}.

I-V curves for the case with slanted coupling are very similar in form, but with lower threshold voltages.

The exact location of the threshold voltages depends on the disorder realisation in the system.
Different experimental systems will have different disorder realisations, and therefore there are a range of possible threshold voltages.
In any numerical simulation one disorder realisation must be used, and therefore no single numerical I-V curve will exhibit exactly the same threshold voltage as a corresponding single experimental I-V curve unless both the theoretical and the experimental systems have exactly the same disorder realisation.
General qualitative trends, however, such as the dependence of the thresholds on the coupling $\alpha$, remain true across different disorder realisations.

\begin{figure}
	\includegraphics[width=1\linewidth]{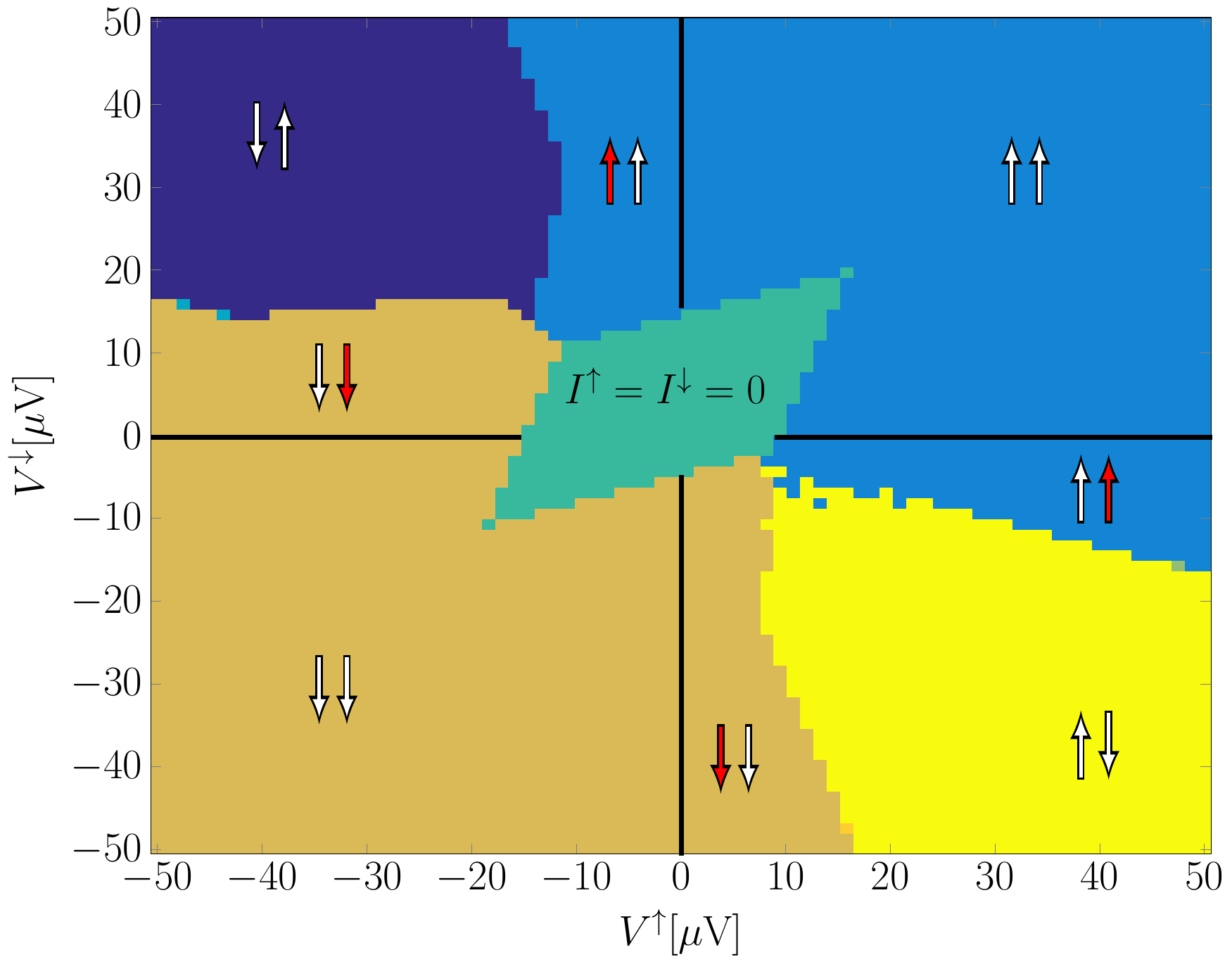}
	\caption{\label{fig:CondDiag} Conduction behaviour of a bilinear Josephson junction array as a function $V^\uparrow$ and $V^\downarrow$, calculated with the same parameters as Fig.~\ref{fig:IVCurves}. Arrows indicate the direction of current (up for positive, down for negative) for the upper and lower arrays respectively. Asymmetry in the diagram is due to asymmetry in the disorder realisation. A region of insulating behaviour is found in the centre of the diagram, with other regions corresponding to currents in both arrays either in the same direction or in opposite directions as indicated by the arrows. Red arrows indicate regions where the current flows in the direction opposite to that of the voltage applied to it, i.e. regions in which Coulomb drag dominates ordinary conduction.}
\end{figure}

This scheme can be easily generalized to the case where the voltage on both arrays is non-zero. 
In this case we obtain an I-V surface, with the current being a function of both $V^\uparrow$ and $V^\downarrow$. 
To study the depinning thresholds and Coulomb drag behaviour change as both voltages are varied, we calculate a conduction diagram, as depicted in Fig.~\ref{fig:CondDiag}. 
Slight asymmetries in the diagram arise due to asymmetry in the disorder realisation.
The conduction diagram for the (physically unrealistic) completely clean bilinear array ($\vec{F} = 0$) has no such asymmetry. 

\begin{figure}
	\includegraphics[width=1\linewidth]{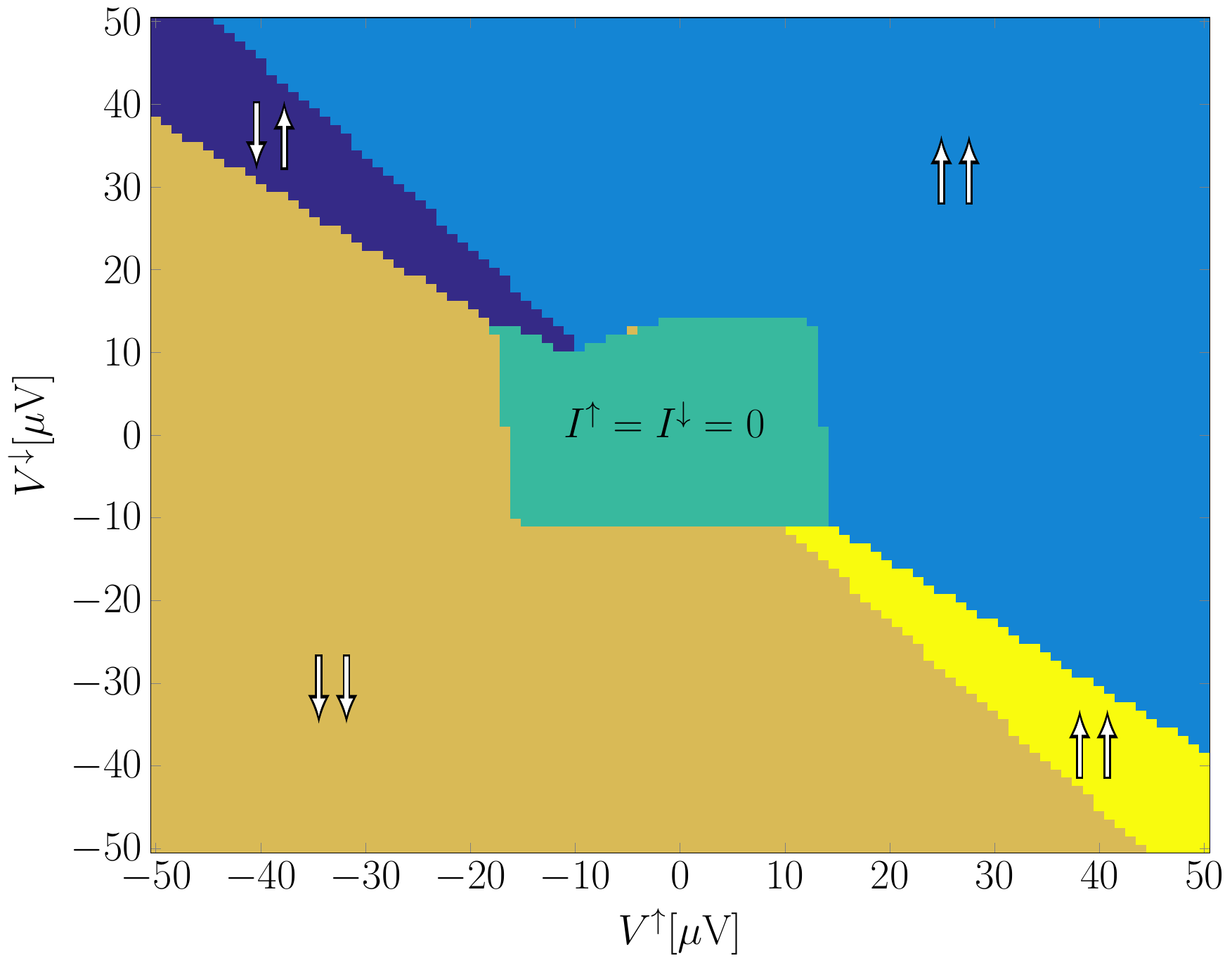}
	\caption{\label{fig:SlantedCD} Conduction diagram for a disordered bilinear Josephson junction array with slanted coupling with the same parameters as \ref{fig:CondDiag}. It can be seen that the regions of overpowering Coulomb drag are much larger than in the straight-coupled array with the same parameters. This can be understood by recognising that the coupling geometry leads to a stronger effective coupling.}
\end{figure}

From the conduction diagram we can see that there exist regions where, in the absence of coupling, a current would flow in one direction, but due to overpowering Coulomb drag it instead flows in the other direction.
These regions of overpowering Coulomb drag are marked in Fig.~\ref{fig:CondDiag} by red arrows.
The conduction diagram for slanted coupling, Fig.~\ref{fig:SlantedCD}, displays even larger regions of overpowering Coulomb drag given the same parameters.
This can be understood as being due to a larger effective coupling between the two arrays with slanted coupling, as in there are more coupling capacitors.

In  numerical simulation, if one wishes to observe the second depinning threshold it is vital that the voltage is applied adiabatically and that the system retains memory of past charge configurations. 
This is to be expected as the microscopic details of depinning theory depend heavily on the existence of metastable states \cite{Brazovskii2003}. 
When the system is driven diabatically we see a current not because the insulating state is unstable against the formation of current, but because the system has been driven sufficiently fast to push it away from stable configurations.
Simulations in which the voltage was increased slowly displayed second threshold behaviour (Fig.~\ref{fig:IVCurves}), but the second threshold was not seen in simulations where the voltage is applied in sharp steps (i.e. diabatically).

Experimentally, to observe the second depinning threshold one should ensure that the this threshold lies below the quasiparticle excitation gap, $V \approx 2N\Delta/e$. The second threshold can be kept at a relatively low voltage if one ensures that the coupling to ground $C_G$ is small, as this ensures that both the dimensionless coupling $\alpha$ and the effective interaction length $\Lambda$ will be large.

One must also ensure that the coupling is not too strong.
Our entire theoretical approach is based around an understanding that the system is in the insulating regime of the quantum phase diagram.
It has been shown that strong coupling between the arrays leads to an insulator-superconductor quantum phase transition in bilinear JJ arrays, even when each array is individually in the insulating state\cite{Choi1998d}.
We therefore only expect our theory to hold when $\sqrt{E_J C_C/(2e)^2} < 2\sqrt{2}/\pi\approx0.9$ for straight coupling and $\sqrt{E_J C_C/(2e)^2} < 2/\pi\approx0.6$ for slanted coupling (as is the case in all of our simulations).

\section{Conclusion}
Bilinear Josephson junction arrays provide an ideal situation in which to study Coulomb drag and depinning in a controllable system. The fact that each array experiences a separate pinning force leads to two depinning thresholds in the I-V curve, and the interplay between depinning and Coulomb drag leads to novel behaviour. Conclusions made with respect to Josephson junction arrays may be applicable to other systems in which both Coulomb drag and depinning effects are present.

I-V curves of bilinear arrays for both straight and slanted coupling display the same threshold behaviour, however both threshold voltages are lower in the case of slanted coupling.
Furthermore, we see that from the conduction diagrams that the array with slanted coupling has much larger regions of overpowering Coulomb drag.

\section{Acknowledgements}
We thank T. Duty and H. Shimada for useful discussions. This work was supported in part by the Australian Research Council under the Discovery and Centre of Excellence funding schemes (project numbers DP140100375 and CE170100039). Computational resources were provided by the NCI National Facility systems at the Australian National University through the National Computational Merit Allocation Scheme supported by the Australian Government.

\bibliography{Bib}

\end{document}